\documentclass[sigconf, 9pt, review=false, screen]{acmart}

\copyrightyear{2025}
\acmYear{2025}
\setcopyright{rightsretained}
\acmConference[GLSVLSI '25]{Great Lakes Symposium on VLSI 2025}{June 30-July 2, 2025}{New Orleans, LA, USA}
\acmBooktitle{Great Lakes Symposium on VLSI 2025 (GLSVLSI '25), June 30-July 2, 2025, New Orleans, LA, USA}
\acmDOI{10.1145/3716368.3735190}
\acmISBN{979-8-4007-1496-2/2025/06}

\usepackage{amsmath,amssymb,amsfonts}
\usepackage{algorithmic}
\usepackage{graphicx}
\usepackage{textcomp}
\usepackage{marvosym} 
\usepackage{ragged2e}

\usepackage{tabularx}
\usepackage{caption}
\usepackage {subcaption}

\usepackage{multirow}
\usepackage{makecell}
\usepackage{booktabs}
\usepackage{diagbox}

\usepackage{CJKutf8}

\usepackage{stfloats}

\def\BibTeX{{\rm B\kern-.05em{\sc i\kern-.025em b}\kern-.08em
    T\kern-.1667em\lower.7ex\hbox{E}\kern-.125emX}}

\def\eg{\textit{e.g.,}\ }
\def\ie{\textit{i.e.,}\ }

\def\etc{\textit{etc.}\ }

\newcommand{\cmd}[1]{\texttt{\small{#1}}}

\usepackage{pifont}       
  
\usepackage{bbding}       
\usepackage{fontawesome}  

\usepackage{etoolbox}
\usepackage{fancyhdr}

\usepackage{textcomp}
\usepackage{pifont}

\begin{document}
\begin{CJK*}{UTF8}{gbsn}




\makeatletter
\renewcommand{\@titlefont}{\sffamily\bfseries\fontsize{15.5}{16}\selectfont}
\makeatother

\title{Bridging the Gap between Hardware Fuzzing and Industrial Verification}







\author{Ruiyang Ma$^{*}$, Tianhao Wei$^{*}$, Jiaxi Zhang, Chun Yang, Jiangfang Yi, Guojie Luo$^{\dagger}$}
\def \authors{Ruiyang Ma, Tianhao Wei, Jiaxi Zhang, Chun Yang, Jiangfang Yi, Guojie Luo}
\renewcommand{\shortauthors}{Ruiyang Ma, Tianhao Wei, Jiaxi Zhang, Chun Yang, Jiangfang Yi, Guojie Luo}
\affiliation{ 
  \institution{\textit{School of Computer Science, Peking University}, Beijing, China\country{}}
  \institution{\textit{National Key Laboratory for Multimedia Information Processing, Peking University}, Beijing, China\country{}}
  \fontsize{9pt}{12pt}\selectfont
  \institution{\{ruiyang, weitianhao\}@stu.pku.edu.cn, \{zhangjiaxi, chunyang, yijiangfang, gluo\}@pku.edu.cn}
}





\begin{abstract}


As hardware design complexity increases, hardware fuzzing emerges as a promising tool for automating the verification process. However, a significant gap still exists before it can be applied in industry. This paper aims to summarize the current progress of hardware fuzzing from an industry-use perspective and propose solutions to bridge the gap between hardware fuzzing and industrial verification.
First, we review recent hardware fuzzing methods and analyze their compatibilities with industrial verification. We establish criteria to assess whether a hardware fuzzing approach is compatible.
Second, we examine whether current verification tools can efficiently support hardware fuzzing. We identify the bottlenecks in hardware fuzzing performance caused by insufficient support from the industrial environment.
To overcome the bottlenecks, we propose a prototype, HwFuzzEnv, providing the necessary support for hardware fuzzing. With this prototype, the previous hardware fuzzing method can achieve a several hundred times speedup in industrial settings.
Our work could serve as a reference for EDA companies, encouraging them to enhance their tools to support hardware fuzzing efficiently in industrial verification.

\end{abstract}



\maketitle

\let\thefootnote\relax\footnotetext{
*Both authors contributed equally to this research. \\
$^{\dagger}$Corresponding author.
}

\section{Introduction}
The growing complexity of hardware designs becomes a significant challenge for verification engineers. It is increasingly difficult for them to identify potential vulnerabilities during the hardware development process~\cite{rfuzz, directfuzz}. Extensive effort is dedicated to creating test cases of the design under test (DUT), which is crucial for achieving DUT coverage closure and discovering bugs in corner cases.

Hardware fuzzing has recently emerged as a promising technique
for reducing the substantial human effort required in hardware verification~\cite{rfuzz,hw-fuzz,directfuzz,difuzz, processorfuzz, thehuzz, noc-fuzz, rtlfuzzlab, genfuzz, socfuzzer}. This innovative approach employs a gray-box test generation strategy, using coverage as feedback to heuristically mutate previous high-quality inputs and generate new stimuli. It has demonstrated a remarkable ability to automatically complete test plans, achieving both code and functional coverage closures~\cite{rfuzz, directfuzz, hw-fuzz}. Furthermore, fuzzing methods have enabled researchers to uncover previously hidden vulnerabilities~\cite{thehuzz, difuzz, processorfuzz}, some even within industrial hardware designs~\cite{c910_bug}. There is significant potential to apply hardware fuzzing in real-world industrial hardware development.

\begin{figure}[t]
\vspace{+5pt}
\centering
\includegraphics[width=0.42\textwidth]{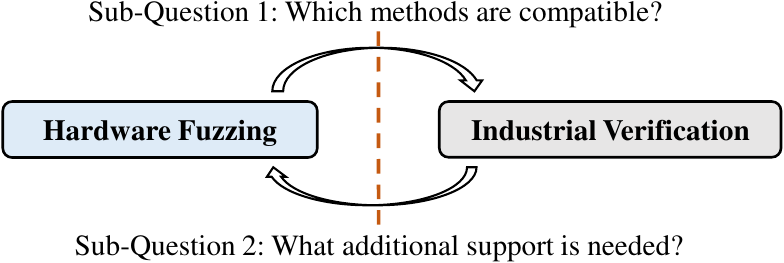}
\vspace{-10pt}
\caption{Our sub-questions about the current gap between hardware fuzzing and industrial verification.}
\vspace{-15pt}
\label{fig:questions}
\end{figure}

However, a key question remains: 
What is the current gap between hardware fuzzing and its practical application in industrial verification?
In this paper, we aim to address this topic by summarizing and exploring two sub-questions, as illustrated in Figure~\ref{fig:questions}.
\begin{itemize}
    \item What types of hardware fuzzing methods are compatible with industrial verification?
    \item What additional support is needed from the industrial verification environment to facilitate hardware fuzzing?
\end{itemize}

For the first sub-question, we review recent influential research on hardware fuzzing and find that many of these works are actually not compatible with industrial verification. 
We analyze their compatibilities from three dimensions: simulation platform, coverage metric, and mutation strategy, which will be detailed in Section~\ref{sec:three}. 
In each dimension, we identify the most appropriate approach for industry.
Firstly, the framework should utilize industrial simulators rather than acceleration platforms.
Secondly, it should focus on achieving traditional coverage metrics that align with real-world test plans. 
Finally, it should eliminate the effort of creating design-specific mutation strategies for each individual design.

After summarizing the first sub-question, we turn to the second: Can current industrial hardware verification tools efficiently support hardware fuzzing?
We select a hardware fuzzing work that aligns with our proposed standards and conduct a quantitative analysis, which will be detailed in Section~\ref{sec:analysis}. Our findings indicate that existing industrial verification tools are not adequately equipped to support hardware fuzzing, resulting in significant inefficiencies. In fact, over 90\% of the performance loss is attributed to the lack of necessary support from the industrial verification environment.

We identify three key limitations in current industrial tools that hinder the efficiency of hardware fuzzing in practical applications.
Firstly, existing industrial simulators lack support for efficient coverage collection. In fuzzing iterations, coverage is collected frequently as feedback and an inefficient coverage interface wastes a large portion of time.
Secondly, industrial tools do not facilitate efficient mutation operations. Current hardware fuzzing works mainly rely on software fuzzers, which are time-consuming due to complex mutation processes and high communication overheads.
Thirdly, industrial tools do not incorporate efficient parallel mechanisms for hardware fuzzing, leaving significant room for optimization.

To overcome the limitations listed above, we develop a prototype hardware fuzzing environment called \textsc{HwFuzzEnv}, with three simple but effective enhancements.
\textsc{HwFuzzEnv} incorporates essential support for hardware fuzzing within the hardware verification environment and associated tools. The experiments demonstrate that our enhancements provide a significant speedup for previous hardware fuzzing practices in industrial applications.

In summary, our main contributions are as follows:


\begin{itemize}
\item We assess the compatibilities of current hardware fuzzing methods for industrial verification and summarize our criteria that determine their suitabilities for industry use.
\item We identify the limitations in industrial tools that do not support hardware fuzzing efficiently. To the best of our knowledge, we are the first to address the lack of support for hardware fuzzing in industrial verification environments.
\item We develop a prototype industrial hardware fuzzing environment named \textsc{HwFuzzEnv}\textsuperscript{1}, which incorporates simple but effective enhancements for each stage of hardware fuzzing, including coverage collection, mutation, and simulation.
\item We evaluate the efficiency and utility of \textsc{HwFuzzEnv}. With the support of \textsc{HwFuzzEnv}, \textsc{RtlFuzzLab}~\cite{rtlfuzzlab} achieves a speedup of up to $21.5\times$ (on average $12.7\times$) with a single thread, and up to $621\times$ (on average $361\times$) with 64 threads.
\end{itemize}

\vspace{-5pt}
\let\thefootnote\relax\footnotetext{
\textsuperscript{1}The project is available at \url{https://github.com/magicYang1573/fast-hw-fuzz}.
}

\section{Background}
\subsection{Industrial Hardware Verification}


In the industrial hardware development lifecycle~~\cite{bg_hw_verify2, bg_hw_verify3}, engineers use hardware description languages (HDLs) at the register-transfer level (RTL) to describe microarchitectural modules. Electronic design automation (EDA) tools synthesize these RTL modules into gate-level designs, which are then mapped to the transistor level and eventually to physical layout for manufacturing.

During hardware development, writing HDL code at the RTL is error-prone. As a result, an overwhelming 60\% of a project's total effort is dedicated to verification~~\cite{siemens_study}. 
Simulation-based verification is a key technique in industrial hardware verification, which necessitates writing input test cases and uses hardware simulators like Synopsys VCS~~\cite{vcs} or Siemens Questa~~\cite{questa} to compare the design's outputs with expected results. 
Hardware coverage is crucial in this process, as it indicates how thoroughly the design has been tested. 

When striving to achieve code and functional coverage closures, random testing fails to reach numerous corner cases within a limited time. Manual test case creation, while precise, is labor-intensive and inefficient. As a result, the research community has sought to automate this process through coverage-directed test generation methods~~\cite{hw_test_generation_survey}. Among these, hardware fuzzing has emerged as a particularly promising technique.

\vspace{-5pt}
\subsection{Hardware Fuzzing}

Fuzzing, originally a concept derived from software testing, denotes the process of testing with minimal or no internal knowledge during the verification phase~~\cite{bg_sw_fuzz_survey}. 
Among the various fuzzing techniques, coverage-guided mutational fuzzing stands out for effectively identifying correctness bugs and security vulnerabilities~~\cite{afl,libfuzzer}.
Recently, researchers have tried to apply the concepts behind software fuzzing to the hardware verification domain.

\begin{figure}[t]
\centering
\includegraphics[width=0.43\textwidth]{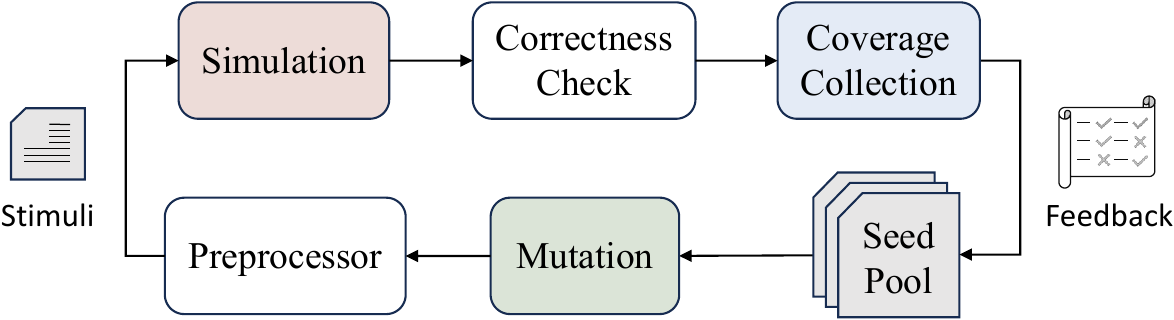}
\vspace{-10pt}
\caption{Basic hardware fuzzing workflow.}
\vspace{-15pt}
\label{fig:fuzz-flow}
\end{figure}

Figure~\ref{fig:fuzz-flow} shows the basic workflow of hardware fuzzing.
The fuzzing process starts with a selection from the seed pool. The seed undergoes mutation to generate new inputs, which are then translated into DUT-compatible stimuli by a preprocessor. The HDL simulator runs the simulation and collects coverage data as feedback. Inputs revealing new hardware states (\ie new coverpoints) are retained as seeds for future fuzzing iterations. This process repeats until no new states emerge for a set number of iterations. Throughout, assertions or golden-model cross-checkings ensure the DUT's correct behaviors. 
In the following section, we categorize existing hardware fuzzing studies and evaluate their compatibilities with the industrial hardware verification environment.



\vspace{-5pt}
\section{Industry-Compatible Hardware Fuzzing}
\label{sec:three}
\begin{figure*}[t]
\centering
\includegraphics[width=0.75\textwidth]{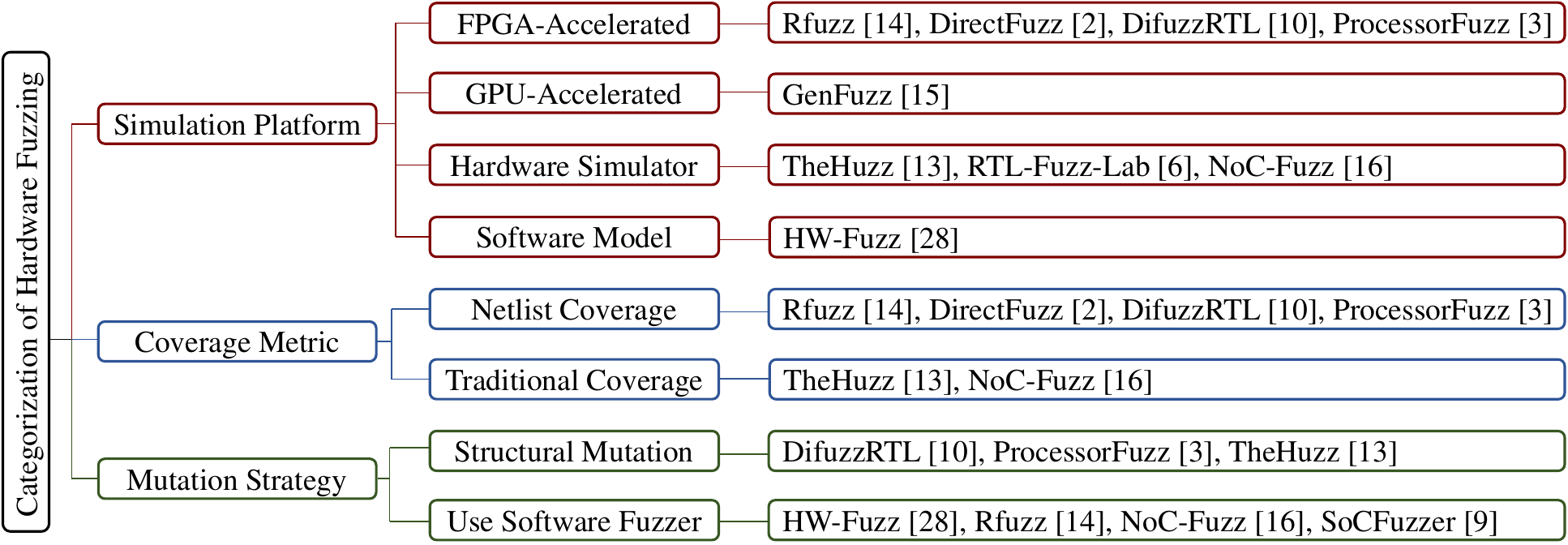}
\vspace{-10pt}
\captionsetup{belowskip=-10pt}\caption{A taxonomy of hardware fuzzing studies with representative examples.}
\label{fig:categorization}
\end{figure*}

The workflow of hardware fuzzing consists of three main parts: a HDL simulation platform for correctness checking, a coverage mechanism for feedback collecting, and a mutation engine for input space exploring, as shown in Figure~\ref{fig:fuzz-flow}. 
Based on the different implementation strategies of each stage, we categorize existing fuzzing research from three dimensions, as depicted in Figure~\ref{fig:categorization}. Each type of methods is evaluated for its advantages and potential limitations in the context of industrial verification.




\vspace{-5pt}
\subsection{Simulation Platform}
In hardware fuzzing, a simulation platform is essential for observing hardware behaviors and collecting coverage metrics. Traditional hardware simulators, such as Synopsys VCS~\cite{vcs} and Siemens Questa~\cite{questa}, often require significant time to run simulations, particularly when dealing with complex hardware designs.

To accelerate the fuzzing process, some studies use FPGA platforms for hardware simulation~\cite{rfuzz, directfuzz, difuzz}. This approach necessitates substantial modifications to normal fuzzing operations. One challenge is that FPGA implementations can only collect coverage data at the netlist level, which is often incompatible with existing industrial test plans. Besides, utilizing FPGAs efficiently poses technical difficulties for industrial verification engineers, particularly in terms of the communication between the CPU-based fuzzer and the FPGA-based simulator. 
Similar challenges also exist when using GPUs for simulation acceleration~\cite{genfuzz}, making these acceleration platforms less suitable for industrial hardware verification.

An alternative approach involves fuzzing hardware in the software domain by translating hardware RTL designs into software models using open-source tools like Verilator~\cite{verilator}. It allows for the application of software fuzzers~\cite{hw-fuzz} to directly verify the RTL software model. However, this method faces challenges in ensuring the software model's equivalence to the original hardware design, which is unacceptable in industrial verification~\cite{bg_hw_fuzz_survey}. 

Therefore, in industrial hardware verification, using a standard hardware simulator as the simulation platform is necessary, even though it may not provide the fastest execution speed. Such simulators offer mature support for coverage collection, making hardware fuzzing compatible with existing industrial workflows.

\subsection{Coverage Metric}
Hardware fuzzing is primarily guided by coverage. Initially, netlist-level coverage metrics are the focus of hardware fuzzing studies~\cite{rfuzz, directfuzz, difuzz, processorfuzz}. For example, \textsc{RFuzz}~\cite{rfuzz} introduces muxiplexer coverage, treating the select signal of each 2:1 multiplexer as a distinct coverpoint.
\textsc{DifuzzRTL}~\cite{difuzz} employs register coverage by using the states of CPU control registers as coverpoints.

However, industry engineers often show limited interest in these novel coverage types, as they are not included in the test plans~\cite{bg_hw_fuzz_survey, thehuzz}. They prefer verification methods and coverage metrics for RTL, the same level at which hardware designs are written. Moreover, translating these new coverage types into traditional metrics is challenging. It is hard to determine whether a line of code has been thoroughly tested based on multiplexer or register coverage alone.

To bridge this gap, some studies have developed hardware fuzzing frameworks that align with traditional coverage metrics~\cite{thehuzz, noc-fuzz}. This makes hardware fuzzing more compatible with industrial verification plans and also promotes the use of industrial hardware simulators in hardware fuzzing workflows.

\subsection{Mutation Strategy}
Mutation strategy is crucial for efficient space exploration in hardware fuzzing. By mutating high-quality seed inputs, newly generated test cases progressively uncover hardware behaviors. Typically, hardware designs necessitate valid stimuli that adhere to specific protocols or standards, indicating that mutation engines must produce diverse inputs while ensuring their compliance with the design's interface and functionality.

Some studies implement structured and grammar-aware mutation strategies tailored to the input format~\cite{difuzz, thehuzz, processorfuzz}. 
For instance, when fuzzing a CPU, the mutation process is customized for the specific instruction set architecture, generating valid CPU instruction sequences. However, this approach is overcomplicated for verification engineers, which requires considerations of the instruction formats as well as the evolutionary algorithms. Given that developing such a complex mutation engine is not an ordinary skill for industrial engineers, and considering the need of unique mutation strategies for each design, it is impractical to apply structural mutation-based fuzzing to industrial verification workflows.

A more accessible and general solution is to use mature software fuzzing tools~\cite{afl,libfuzzer} as mutation engines~\cite{rfuzz, hw-fuzz, rtlfuzzlab, socfuzzer, hypfuzz}. This approach requires an additional translation step to convert the generated inputs from software fuzzers into valid hardware stimuli. For instance, AFL~\cite{afl} generates input files in a binary format. To produce valid stimuli, a hardware fuzzing grammar is designed to map the binary streams to specific hardware transactions. This allows fuzzer to use the same mutation flow across different hardware designs. Consequently, verification engineers only need to focus on the mapping strategies, while mature software fuzzers provide powerful heuristic exploration capabilities, rendering hardware fuzzing a more user-friendly method for industrial verification.

In summary, we categorize existing hardware fuzzing studies from three dimensions and analyze their suitabilities for integration into industrial hardware verification workflows. We outline three criteria to help identify hardware fuzzing methods that are compatible with existing industrial verification:
\begin{itemize}
\item Firstly, it should use industrial hardware simulators rather than acceleration platforms with FPGAs or GPUs.
\item Secondly, it should aim at achieving traditional coverage metrics which are included in industrial test plans.
\item Thirdly, it should use a grammar-agnostic mutation strategy to avoid the need of design-specific structural mutations.
\end{itemize}

\section{Analysis of Industrial Gaps}
\label{sec:analysis}


In this section, we investigate whether existing industrial hardware verification tools efficiently support hardware fuzzing. If not, what additional support is needed?
We use \textsc{RtlFuzzLab}~\cite{rtlfuzzlab} as a case study. 
This open-source framework meets our defined industrial criteria well and implements two of the most influential hardware fuzzing algorithms~\cite{rfuzz, hw-fuzz} in recent years, offering representative and referential significance.
Our analysis below evaluates its performance from a time perspective, and we aim to identify the reasons for any inefficiencies observed.

\vspace{-5pt}
\subsection{Time Distribution of Hardware Fuzzing}
\label{subsec:time}
We evaluate the performance of the hardware fuzzing framework on two designs of different scales: a TLI2C IP comprising several thousand lines of code~\cite{sifive} and the Rocket Core with over 40,000 lines of code~\cite{rocket}.
In hardware fuzzing, the initial seed is crucial, as it affects the generation of new inputs, the simulation time for each fuzzing iteration, and the overall frequency of fuzzing.
Therefore, we conduct our experiments with initial seeds of varying lengths.

\begin{figure}[t]
    \centering
    \begin{subfigure}{0.43\textwidth}
        \includegraphics[width=\textwidth]{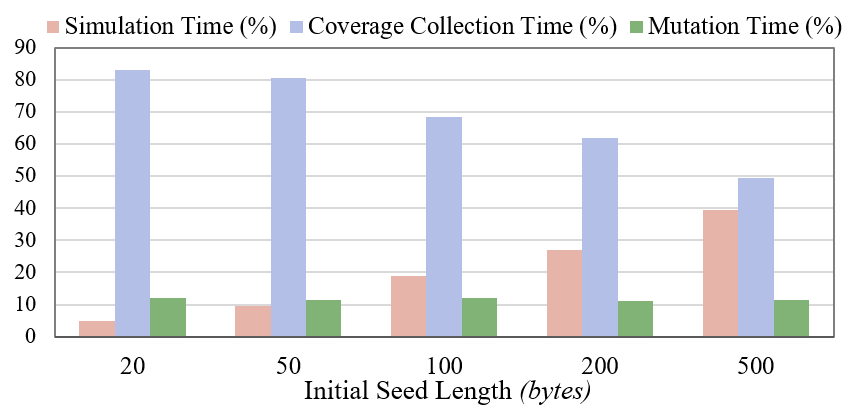}
        \captionsetup{skip=-1pt}
        \caption{I2C}
        \vspace{-2pt}
    \end{subfigure}
    \begin{subfigure}{0.43\textwidth}
        \includegraphics[width=\textwidth]{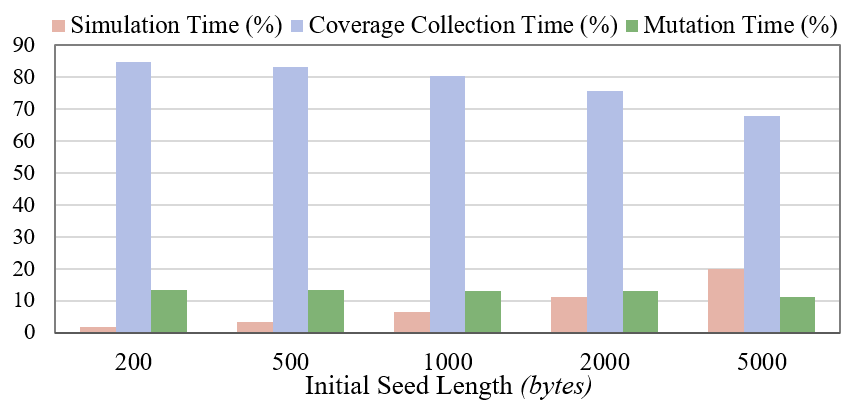}
        \captionsetup{skip=-1pt}
        \caption{Rocket Core}
    \end{subfigure}
    \vspace{-12pt}
    \caption{Time distribution across hardware fuzzing stages.}
    \vspace{-15pt}
    \label{fig:time-cost}
\end{figure}

As illustrated in Section~\ref{sec:three}, hardware fuzzing mainly consists of three stages: coverage collection, mutation, and simulation, with their time distribution shown in Figure~\ref{fig:time-cost}. 
Counter-intuitively, the coverage collection and mutation stages occupy a significant portion of the total time, exceeding 90\% when the initial seed length is small. As the seed length increases, the simulation time grows, while the proportion of time spent on coverage collection and mutation remains over 50\%. The results indicate that existing industrial verification tools are not adequately equipped to support hardware fuzzing, resulting in significant time inefficiencies. 



\vspace{-5pt}
\subsection{Industrial Gaps for Hardware Fuzzing}

\textit{Coverage Collection Inefficiency.}
Existing fuzzing works use Synopsys VCS~\cite{vcs} or Verilator~\cite{verilator} to collect coverage information at simulation runtime. 
VCS maintains a structured and hierarchical coverage database for engineers to extract coverage reports~\cite{vcs_cov}, while Verilator outputs coverage report in a text format, including each coverpoint's definition and hit count~\cite{verilator}.
Both VCS and Verilator only support a detailed and structured coverage collection mechanism, which is untailored and redundant for hardware fuzzing. Specifically, this mechanism is designed for long-term simulation, providing engineers with comprehensive reports for analysis. In contrast, hardware fuzzing requires frequent coverage updates in each iteration, which necessitates a more sketched and efficient coverage interface provided by industrial simulators. 

\textit{Mutation Inefficiency.}
Many hardware fuzzing studies use mature software fuzzers like AFL~\cite{afl} as their mutation engine~\cite{rfuzz, hw-fuzz, rtlfuzzlab, noc-fuzz}. However, this can slow down the fuzzing process for two reasons. 
Firstly, AFL requires inter-process communications with the hardware simulation environment, which is time-consuming. 
Secondly, AFL utilizes complex mutation operations (\eg bitflip, havoc, trim, splice \etc) to mutate structured software inputs in a grammar-agnostic manner. Given the generally simpler nature of hardware interface protocols and the potential of designing hardware fuzzing grammars~\cite{hw-fuzz} to generate valid stimuli, these mutation operations can be simplified. 
Considering these factors, there is a clear need of an efficient mutation engine for hardware fuzzing, which is seamlessly integrated within the industrial simulation environment.

\textit{Serial Execution Inefficiency.}
As illustrated in~\ref{subsec:time}, the simulation stage only occupies a relatively small fraction of time in the fuzzing process. Therefore, most previous methods are restricted to a single-input fuzzing mechanism that runs one simulation thread at a time, overlooking the potential advantages of batch simulation with multiple threads.
Hardware fuzzing is not strictly serial and can benefit greatly from parallel execution.
After the coverage collection and mutation stages are supported and optimized, we re-examine the entire hardware fuzzing process, 
trying to enable the simulation environment to efficiently support hardware fuzzing parallelism and fully exploit the capabilities of multi-core processors.




\vspace{-5pt}
\section{Hardware Fuzzing Environment Prototype}
\label{sec:five}

\begin{figure}[t]
\centering
\includegraphics[width=0.45\textwidth]{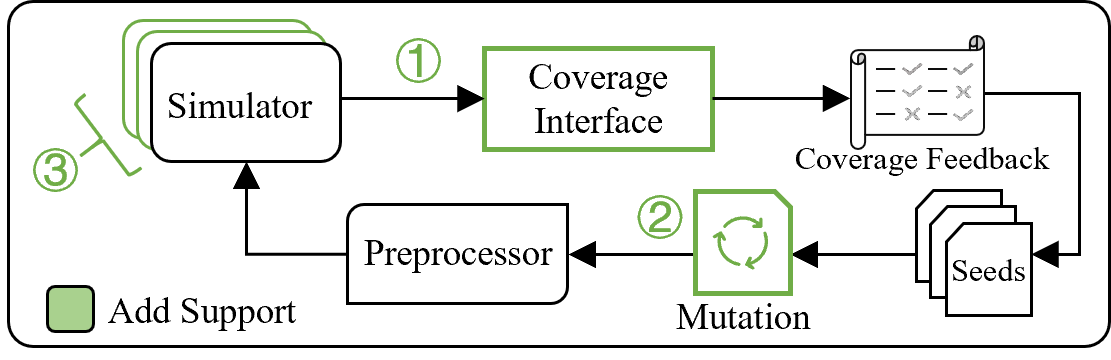}
\vspace{-10pt}
\caption{Overview of our fuzzing environment prototype.}
\vspace{-15pt}
\label{fig:env_overview}
\end{figure}

The analysis in Section~\ref{sec:analysis} shows that current industrial environment cannot adequately support hardware fuzzing, leading to significant time inefficiencies. Therefore, focused on the three inefficiency reasons above, we develop an industrial environment prototype, \textsc{HwFuzzEnv}, which enhances support for hardware fuzzing.
Figure~\ref{fig:env_overview} illustrates the overview of our prototype.
It introduces three simple but effective enhancements.
Firstly, a sketched coverage interface is proposed for the hardware simulator to improve coverage collection efficiency (\ding{192} in Figure~\ref{fig:env_overview}). Secondly, a simplified mutation engine is developed within the hardware simulation environment to support efficient mutation (\ding{193} in Figure~\ref{fig:env_overview}). Lastly, a multi-thread pipelined mechanism is proposed to support efficient parallelism and speedup the whole fuzzing workflow (\ding{194} in Figure~\ref{fig:env_overview}).

\vspace{-5pt}
\subsection{Sketched Coverage Interface}
\label{sec:cov-opt}

Previous analysis reveals that the inefficiency of coverage collection arises from industrial simulators' lack of support for an efficient coverage interface, which is essential for hardware fuzzing.
Generally, hardware fuzzing engines require only a coverage vector (\ie a vector of coverpoints' hit counts) as feedback for heuristic mutation~\cite{rfuzz, hw-fuzz, directfuzz, difuzz, noc-fuzz, thehuzz}. The specific locations and order of coverpoints in the source code do not influence the performance of fuzzing. To support efficient industrial hardware fuzzing, simulators need an interface that can directly acquire such a coverage vector instead of parsing comprehensive coverage reports during fuzzing. Since we do not have access to the source code of commercial simulators, we investigate the open-source Verilator~\cite{verilator} to implement the sketched coverage interface in \textsc{HwFuzzEnv}. 

Verilator instruments coverage monitors in the abstract syntax tree (AST) nodes generated from the hardware design. These monitors trigger coverage counters when specific simulation conditions are met. Each counter is then mapped to an element within a coverage vector, incrementing the corresponding value continuously throughout the runtime. 
We modify a few lines of source code in Verilator to directly access this coverage vector. We have guaranteed that the coverage vector derived from our sketched coverage interface is equivalent to the one obtained from the original coverage interface. The only difference lies in the order of the coverpoints within the vector, which is unnecessary for hardware fuzzing.

\vspace{-5pt}
\subsection{Simplified Mutation Engine}
\label{sec:mutate-opt}

\begin{figure}[t]
\centering
\includegraphics[width=0.42\textwidth]{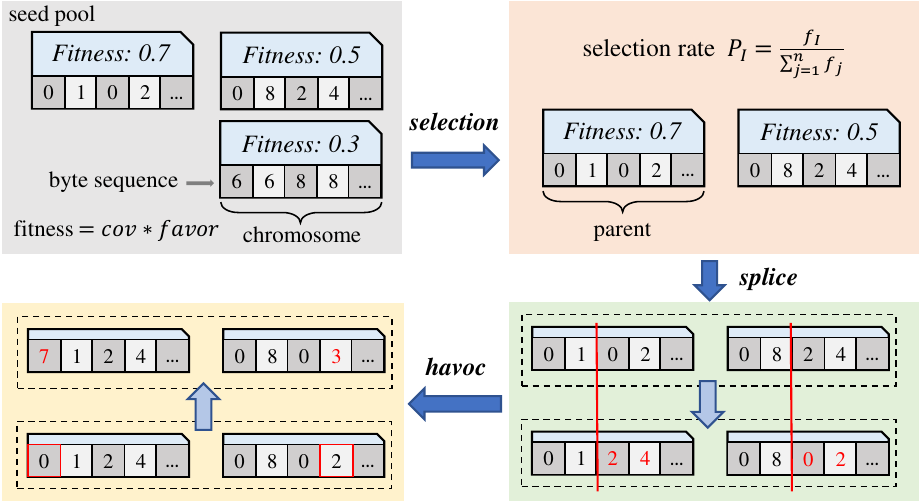}
\vspace{-10pt}
\captionsetup{belowskip=0pt}\caption{Simplified mutation engine in \textsc{HwFuzzEnv}.}
\vspace{-15pt}
\label{fig:mutate-flow}
\end{figure}

According to analysis in Section~\ref{sec:analysis}, it is necessary to integrate a lightweight mutation engine into the hardware simulation environment. This mutation engine should perform effective and efficient mutation operations.
Figure~\ref{fig:mutate-flow} illustrates the simplified mutation engine for hardware fuzzing in \textsc{HwFuzzEnv}.

In the mutation flow, each chromosome is represented as a sequence of bytes, which allows us to define a parametric generator similar to the one proposed by Zest~\cite{zest}. The parametric generator is a function that takes a sequence of untyped parameters and produces a structured input according to the hardware fuzzing grammar. A byte serves as the basic unit of the parametric generator, enhancing the granularity of mutation operations while reducing the time required. 

\textit{Selection:} The selection process prioritizes individuals with higher fitness from the input pool for future mutation. The fitness function takes into account two primary factors. Firstly, it considers the coverage rate achieved by the input, favoring those that achieve more coverage. Secondly, it assesses whether the input is uniquely responsible for reaching certain coverpoints. If a coverpoint can only be covered by a specific input seed, the fitness of the seed is multiplied by a $favor$ factor, increasing its chance for selection.

\textit{Mutation:} 
By carefully examining previous hardware fuzzing works that use software fuzzer AFL~\cite{rfuzz, hw-fuzz, noc-fuzz-tcad} as mutation engine, we find that all the interesting inputs are generated from its havoc and splice operation. Therefore, we only implement these two operations in our simplified mutation engine. The havoc operation randomly modifies byte values in the chromosome, and the splice operation involves a crossover between two selected chromosomes. 

\vspace{-5pt}
\subsection{Simulation Parallelization}
\label{sec:multi-thread-opt}

The simulation workflow is an iterative process that includes the following steps: i) getting input from mutation engine, ii) running HDL simulation, and iii) updating seed corpus using new coverage. 
After the industrial verification environment has been enhanced to support efficient coverage collection and mutation, we can focus on accelerating the simulation step within \textsc{HwFuzzEnv}. To speed up the simulation process, we implement batch simulations by running multiple simulations concurrently across multiple threads.
The framework gets multiple inputs per iteration and dispatches them to individual threads. Each thread operates a hardware simulator in parallel, and the simulation time can be amortized thereby. Once all threads complete the simulation tasks, their coverage information is collected to update the seed corpus in the mutation engine.


The time costs associated with steps i) and iii) are relatively minimal for a single thread. However, as the number of threads increases, these costs escalate due to the multiplication of input acquisitions and coverage updates. Specifically, the time consumed by these two steps begins to approximate that of the simulation stage. 
Given that these steps do not strictly demand serial processing, we utilize a pipelined approach in \textsc{HwFuzzEnv} to improve the throughput of simulation and hardware fuzzing.

\begin{figure}[t]
\centering
\includegraphics[width=0.42\textwidth]{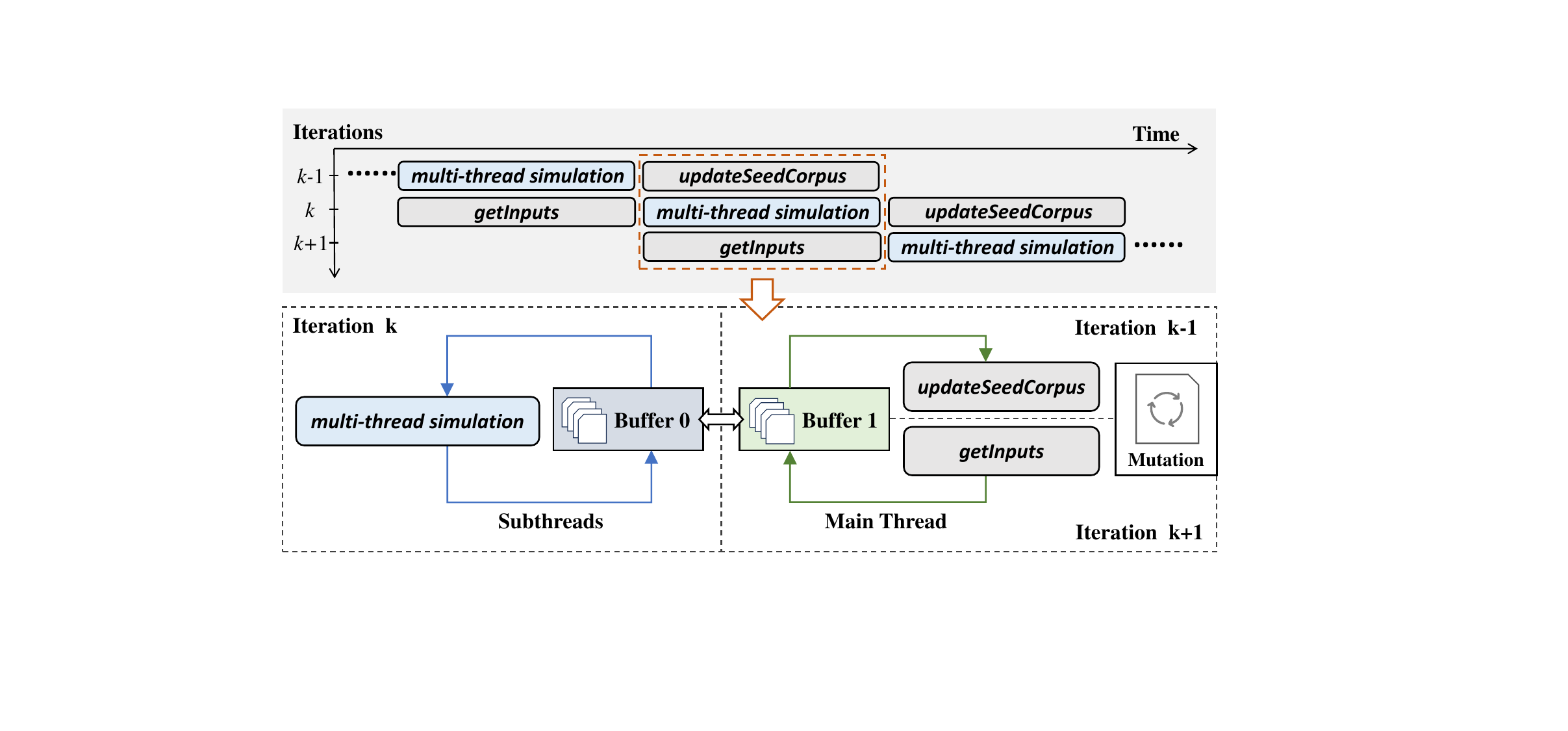}
\vspace{-10pt}
\caption{Multi-thread pipelined mechanism in \textsc{HwFuzzEnv}.}
\vspace{-15pt}
\label{fig:pipeline-multi-thread}
\end{figure}

Figure~\ref{fig:pipeline-multi-thread} illustrates our multi-thread pipelined fuzzing mechanism, in which different parts of three iterations run concurrently at every timestamp. While each child thread is running simulation for iteration $k$, the main thread simultaneously \cmd{updateSeedCorpus} using the coverage of iteration $k{-}1$ and \cmd{getInputs} for iteration $k{+}1$. Notably, the seed corpus for the \cmd{getInputs} function during iteration $k{+}1$ does not include the updated information from iteration $k$ (as it is still in simulation).
To transfer data as a pipeline, we utilize a ping-pong buffer to store the inputs and coverage data. In this manner, we optimize the utilization of multi-threading and speed up hardware fuzzing further. 


\vspace{-5pt}
\section{Experiments and Results}

We evaluate \textsc{HwFuzzEnv} based on the open-source hardware fuzzing framework \textsc{RtlFuzzLab}~\cite{rtlfuzzlab}, which employs AFL~\cite{afl} for mutation and Verilator~\cite{verilator} for simulation. We collect the branch coverage as fuzzing guidance in our experiments and the workflows are the same for other code coverage or functional coverage. 
We comprehensively evaluate on a range of open-source RTL designs, including TileLink Peripheral IPs~\cite{sifive}, RISC-V Sodor Cores~\cite{sodor}, and the RISC-V Rocket Core~\cite{rocket}. A detailed list of the benchmarks and their characteristics is available in Table~\ref{tab:benchmarks_characteristics}. 




We conduct our experiments on a Ubuntu 20.04 system with 384GB RAM and two 3GHz Intel Xeon Gold-6248R CPUs, offering 48 physical cores (96 threads) in total.

\begin{table}[t]
\centering
\caption{Benchmarks and characteristics.}
\vspace{-10pt}
\renewcommand{\arraystretch}{0.85}
\small
\begin{tabular}{|>{\centering}p{0.4cm}c|>{\centering}p{1.6cm}|c|c|}
\hline
Source          & Name          & Input Width & FIRRTL Lines & Cover Points  \\ \hline
                    & I2C         & 165 & 2373  & 245  \\
\cite{sifive}  & PWM         & 163 & 2452  & 181  \\
                  & UART        & 164 & 2416  & 136  \\ \hline
    & Sodor1Stage & 35 & 3617  & 303  \\
  \cite{sodor}    & Sodor3Stage & 35 & 4020  & 314  \\
                  & Sodor5Stage & 35 & 4088  & 341  \\ \hline
 \cite{rocket}  & Rocket Core & 239 & 43856 & 2378 \\ \hline
\end{tabular}
\label{tab:benchmarks_characteristics}
\vspace{-15pt}
\end{table}

\vspace{-5pt}
\subsection{Coverage Interface Comparison}
\label{sec:cov-exp}
In Section~\ref{sec:cov-opt}, we implement the sketched coverage interface in Verilator to support efficient coverage collection for hardware fuzzing. A comparative experiment is conducted to evaluate the time efficiency of our sketched coverage interface against the original coverage interface. The fuzzing process is iterated 10,000 times, and the total time spent in coverage collection is shown in Table~\ref{tab:time_comparison}.

The results indicate that, by using the sketched coverage interface of \textsc{HwFuzzEnv}, the coverage collection process achieves a several-hundred-fold speedup. According to the analysis in Section~\ref{subsec:time}, coverage collection is the most time-consuming stage in hardware fuzzing. By implementing the sketched coverage interface, we reduce its time consumption to a negligible amount.

\vspace{-5pt}
\subsection{Mutation Engine Comparison}
\label{sec:mutate-exp}
In Section~\ref{sec:mutate-opt}, we propose a simplified mutation engine in \textsc{HwFuzzEnv}. We conduct a comparative experiment to evaluate its time efficiency against the AFL mutation engine. The fuzzing process is iterated 10,000 times, and the total time spent in mutation is presented in Table~\ref{tab:time_comparison}. The results indicate that our simplified mutation engine achieves a speed dozens of times faster than AFL. 

However, it is also crucial to investigate whether this reduction in mutation operations adversely affects the quality of the generated inputs. We compare the number of fuzzing iterations required to achieve $K$\% branch coverage using our mutation engine and AFL, starting with the same initial seeds. For different designs, the value of $K$ is different because the maximum coverage that can be achieved is different. As demonstrated in Table~\ref{tab:mutation_comparison}, our simplified mutation engine attains equivalent coverage with much fewer fuzzing iterations. The results suggest that our mutation engine improves not only its speed but also the quality of the generated inputs.

\vspace{-5pt}
\subsection{Fuzzing Speed Comparison}
\label{sec:single-time-exp}
\begin{figure}[t]
    \centering
    \begin{subfigure}{0.22\textwidth}
        \includegraphics[width=\textwidth]{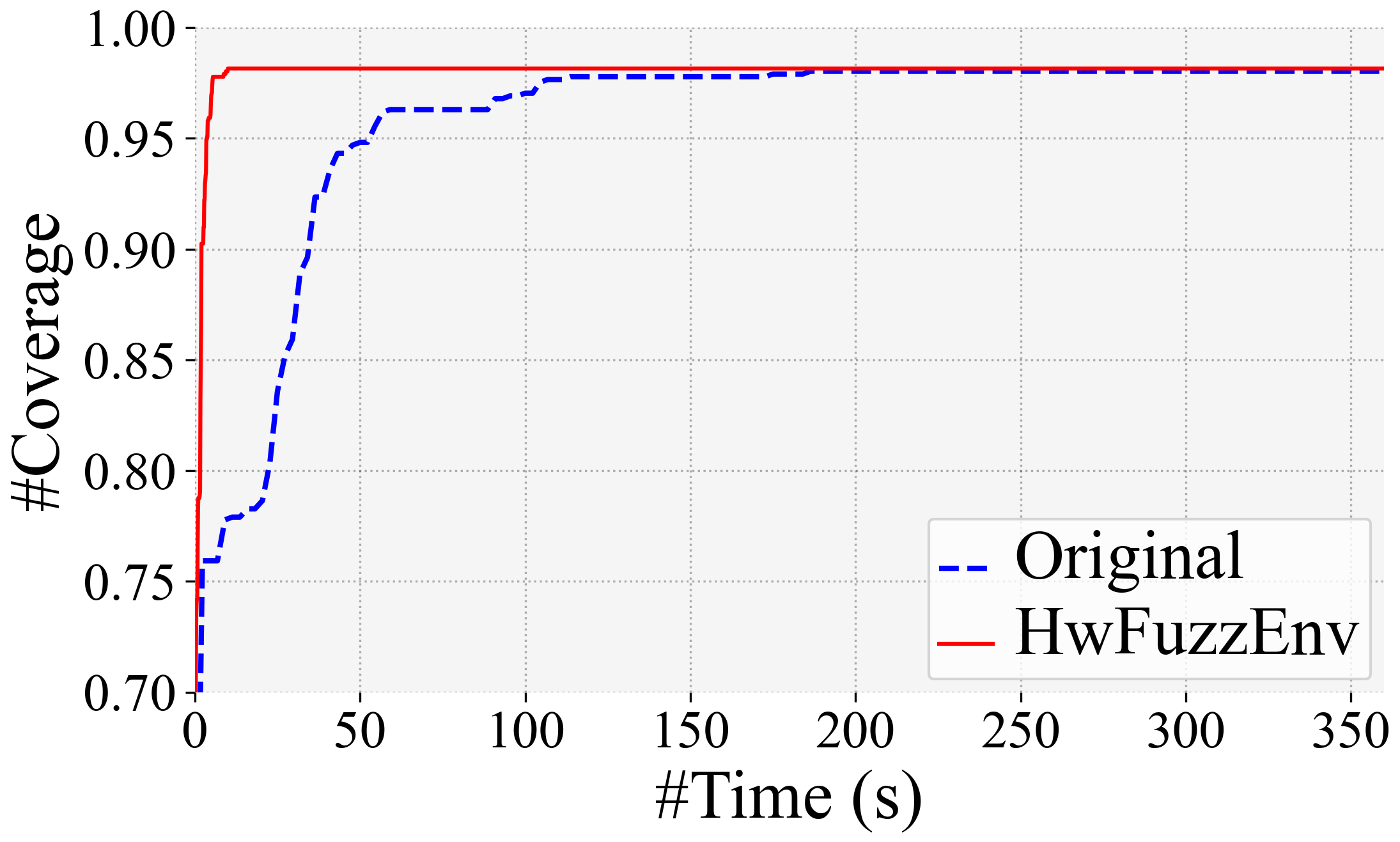}
        \captionsetup{skip=0pt}
        \caption{I2C}
    \end{subfigure}
    \begin{subfigure}{0.22\textwidth}
        \includegraphics[width=\textwidth]{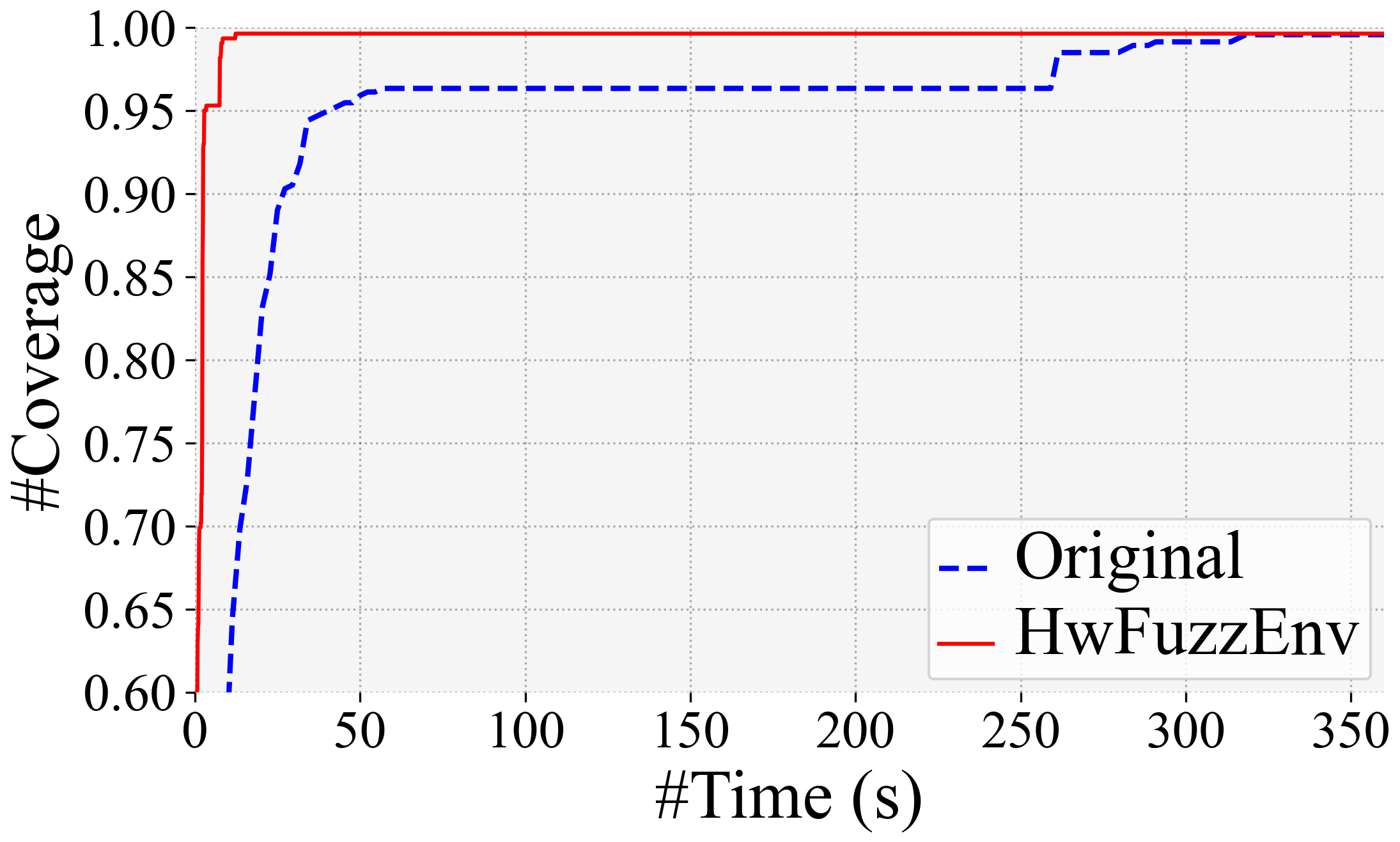}
        \captionsetup{skip=0pt}
        \caption{PWM}
    \end{subfigure}
    \begin{subfigure}{0.22\textwidth}
        \includegraphics[width=\textwidth]{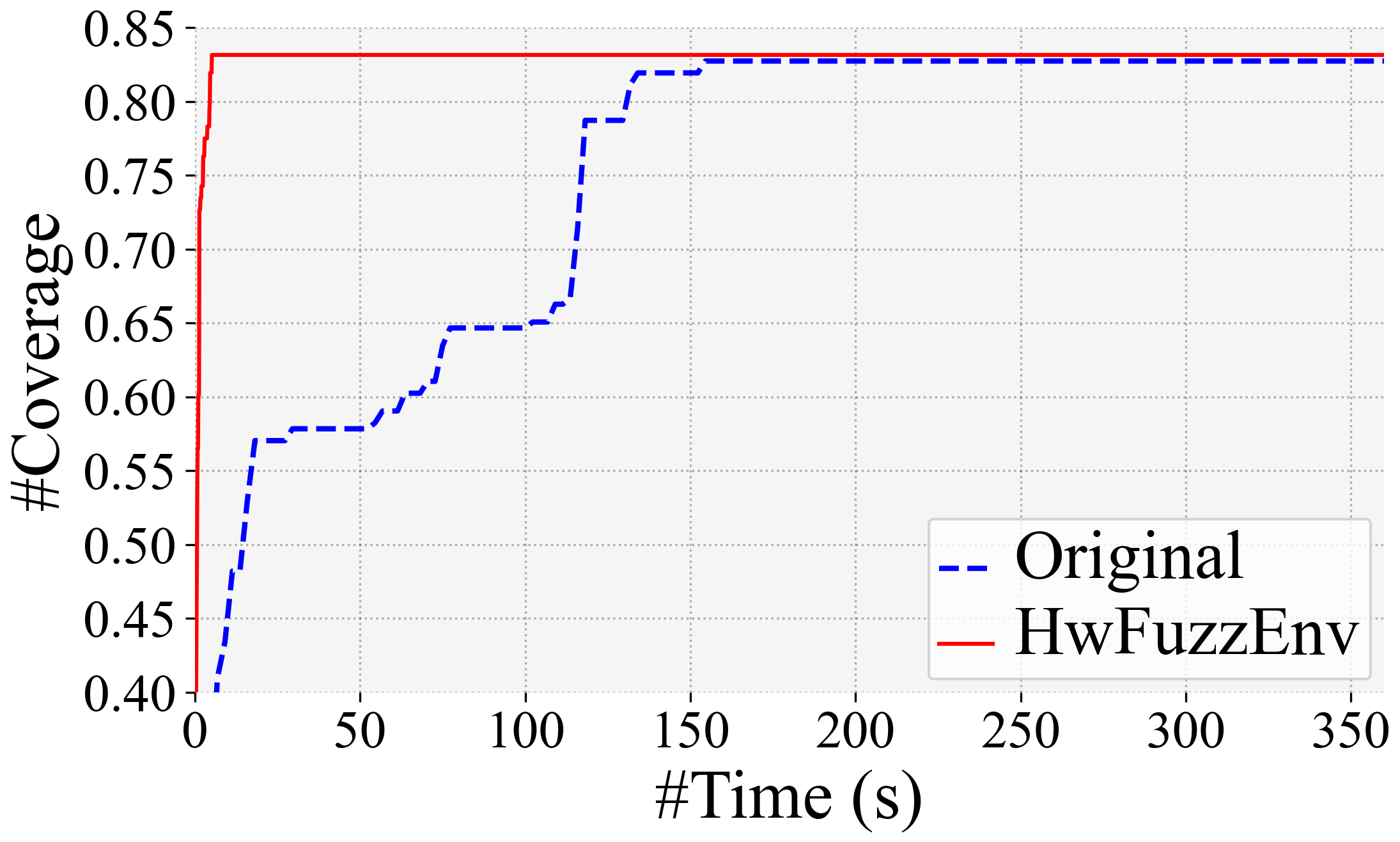}
        \captionsetup{skip=0pt}
        \caption{UART}
    \end{subfigure}
    \begin{subfigure}{0.22\textwidth}
        \includegraphics[width=\textwidth]{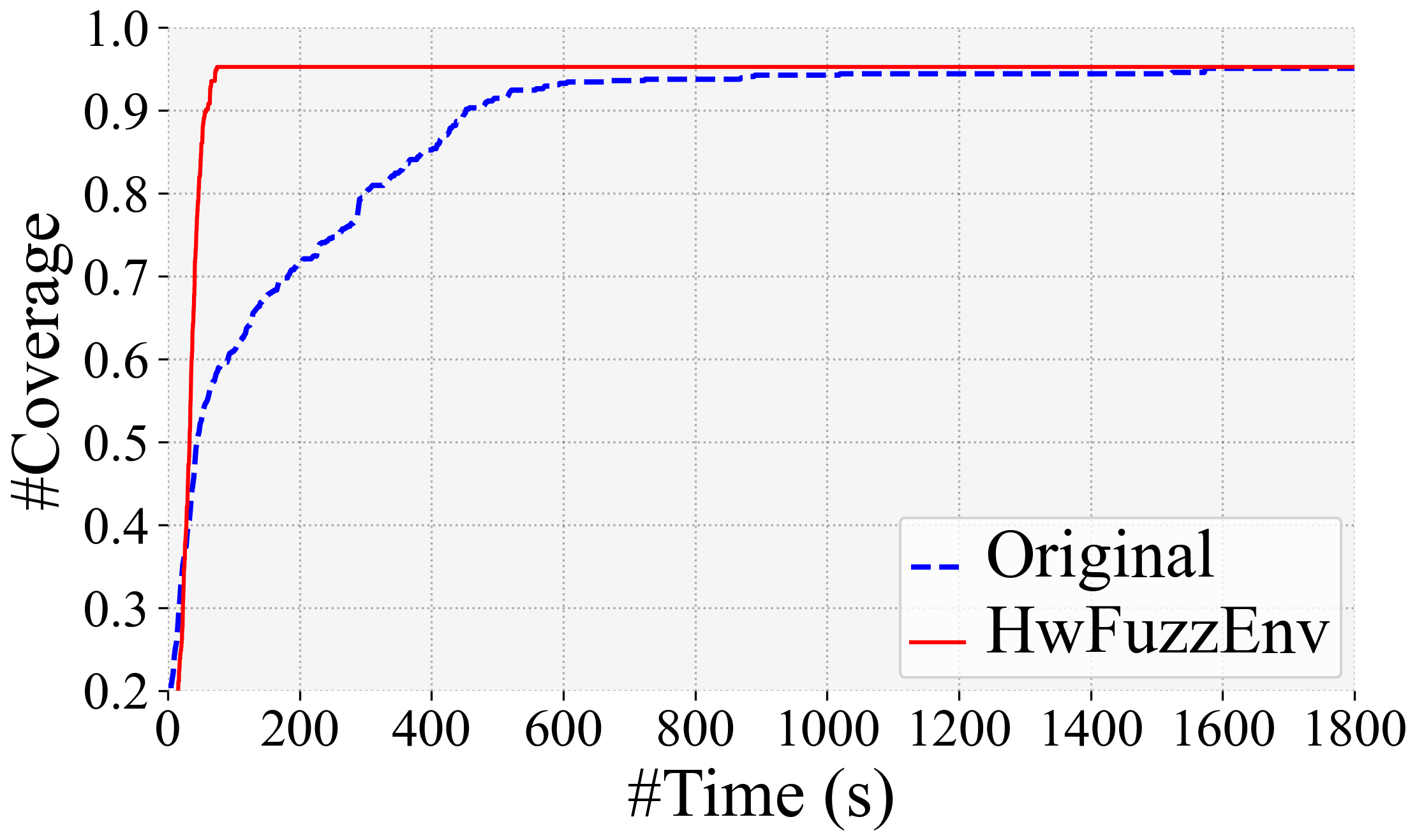}
        \captionsetup{skip=0pt}
        \caption{Sodor 1 Stage}
    \end{subfigure}
    \begin{subfigure}{0.22\textwidth}
        \includegraphics[width=\textwidth]{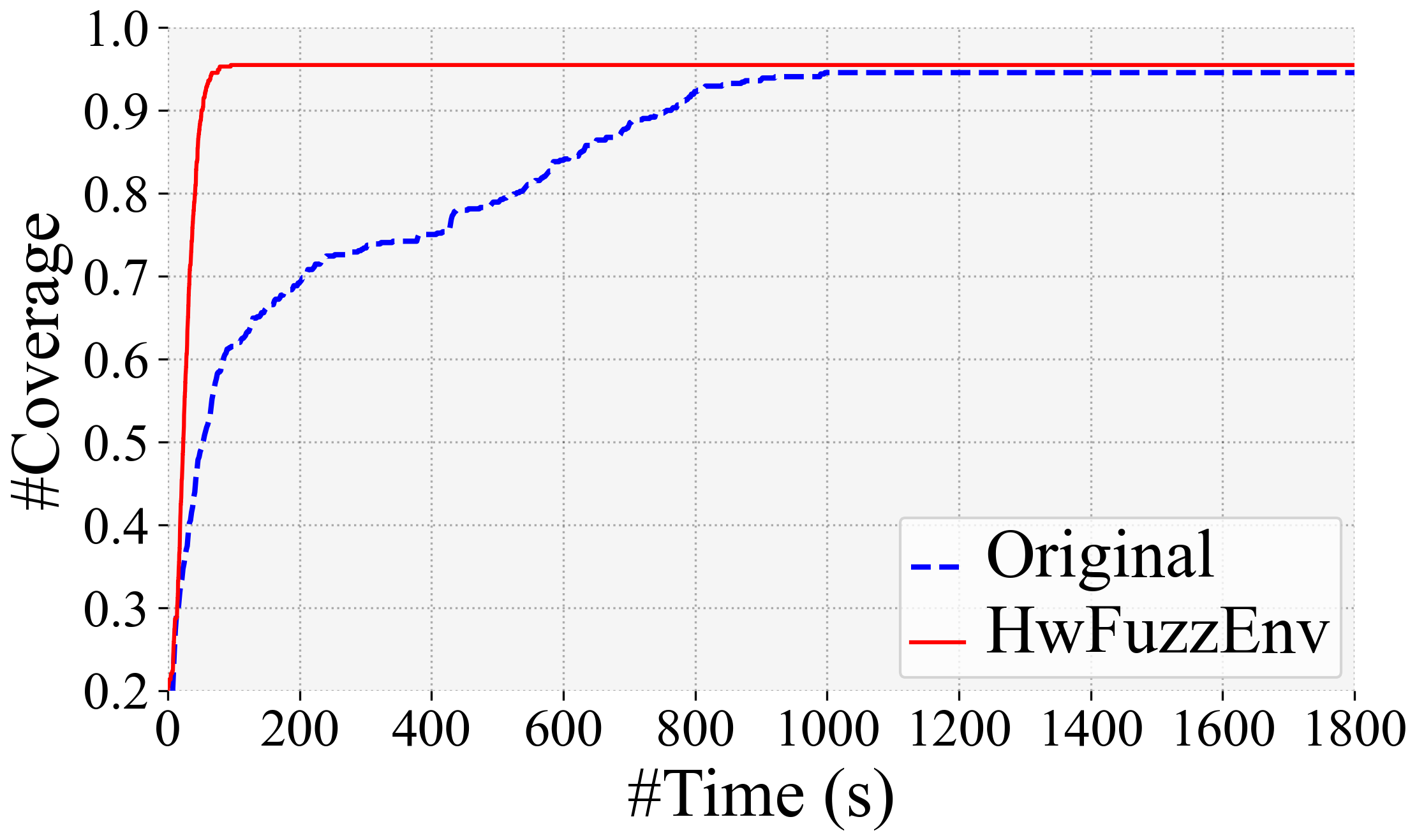}
        \captionsetup{skip=0pt}
        \caption{Sodor 3 Stage}
    \end{subfigure}
    \begin{subfigure}{0.22\textwidth}
        \includegraphics[width=\textwidth]{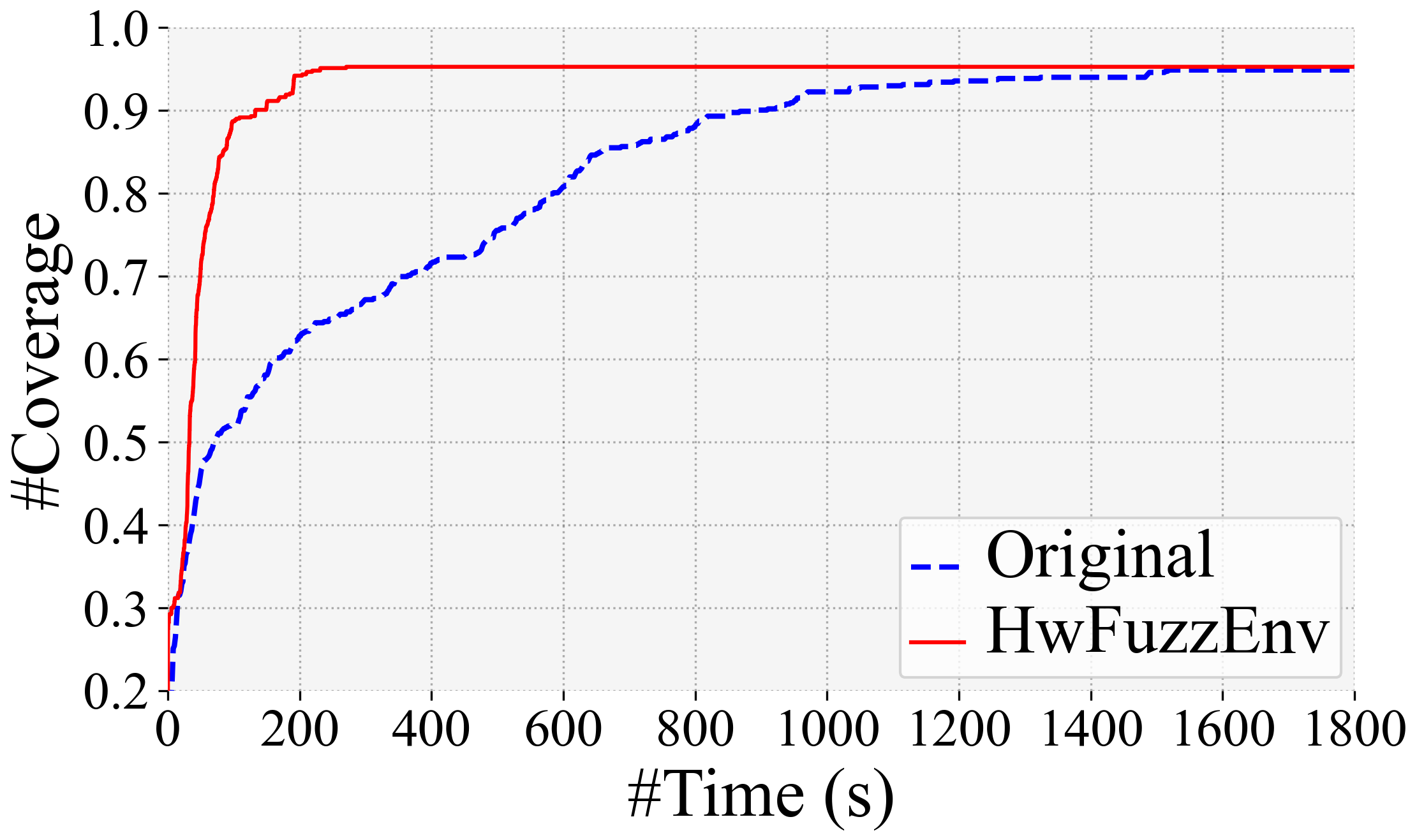}
        \captionsetup{skip=0pt}
        \caption{Sodor 5 Stage}
    \end{subfigure}
    \vspace{-10pt}
    \caption{Comparison on single-thread performance of \textsc{RtlFuzzLab} with and without \textsc{HwFuzzEnv}'s support.}
    \vspace{-15pt}
    \label{fig:single-time-exp}
\end{figure}

After evaluating the efficiency of the sketched coverage interface and simplified mutation engine, we integrate these components and evaluate their combined performance improvement for \textsc{RtlFuzzLab}. In this subsection, we adopt a single-thread fuzzing workflow. The experimental results are presented in Figure~\ref{fig:single-time-exp}.

The results indicate that, with the support of \textsc{HwFuzzEnv}, by optimizing the coverage collection and mutation stage, \textsc{RtlFuzzLab} improves the fuzzing speed by $6.2\times$-$21.5\times$ (on average $12.7\times$). Remarkably, this substantial improvement is achieved with only a single thread. In the following subsection, we extend our investigation to include multi-thread fuzzing experiments.


\begin{table}[t]
\centering
\caption{Comparison on coverage collection and mutation efficiency of \textsc{RtlFuzzLab} with and without \textsc{HwFuzzEnv}'s support.}
\vspace{-10pt}
\renewcommand{\arraystretch}{0.90}
\small
\begin{tabular}{|c|cc|cc|}
\hline
\multirow{2}{*}{\centering Benchmark} & \multicolumn{2}{c|}{Time on Coverage (s)} & \multicolumn{2}{c|}{Time on Mutation (s)} \\ \cline{2-5} 
         & Original         & Sketched        & Original  & Simplified \\ \hline
I2C         & 41.61  & 0.09 (\textbf{462}$\times$)  & 17.20 & 0.37 (\textbf{46}$\times$) \\
PWM         & 34.85  & 0.08 (\textbf{435}$\times$)  & 16.73 & 0.41 (\textbf{40}$\times$) \\
UART       & 31.73  & 0.07 (\textbf{453}$\times$)  & 16.88 & 0.35 (\textbf{48}$\times$) \\ \hline
Sodor1Stage  & 51.53  & 0.05 (\textbf{1030}$\times$) & 17.86 & 0.30 (\textbf{59}$\times$) \\
Sodor3Stage  & 53.27  & 0.05 (\textbf{1024}$\times$) & 18.16 & 0.30 (\textbf{60}$\times$) \\
Sodor5Stage & 55.82  & 0.05 (\textbf{1053}$\times$) & 18.06 & 0.31 (\textbf{58}$\times$) \\ \hline
Rocket Core  & 262.73 & 0.26 (\textbf{1010}$\times$) & 21.43 & 0.81 (\textbf{26}$\times$) \\ \hline
\end{tabular}
\label{tab:time_comparison}
\vspace{-15pt}
\end{table}


\begin{table}[t]
\centering
\caption{Comparison on mutation quality of AFL's mutation and \textsc{HwFuzzEnv}'s simplified mutation.}
\vspace{-10pt}
\renewcommand{\arraystretch}{0.85}
\small
\begin{tabular}{|c|>{\centering}p{2cm}c|}
\hline
\multirow{2}{*}{\centering Benchmark} & \multicolumn{2}{c|}{Iterations for Achieving $K$\% Coverage} \\ \cline{2-3} 
         & AFL Mutation         & Simplified Mutation         \\ \hline
I2C ($K$=95)  & 13268  & 6663 (\textbf{2.0}$\times$)   \\
PWM ($K$=95) & 69311  & 36757 (\textbf{1.9}$\times$)   \\
UART ($K$=80) & 81291  & 50908 (\textbf{1.6}$\times$)   \\ \hline
Sodor1Stage ($K$=90)  & 104384  & 48301 (\textbf{2.2}$\times$)  \\
Sodor3Stage ($K$=90) & 124100  & 48895 (\textbf{2.5}$\times$)  \\
Sodor5Stage ($K$=90) & 127446  & 76110 (\textbf{1.7}$\times$) \\ \hline
\end{tabular}
\label{tab:mutation_comparison}
\vspace{-10pt}
\end{table}

\vspace{-5pt}
\subsection{Multi-Thread Comparison}
\label{sec:exp-multi-thread}
We develop a multi-thread pipelined mechanism in \textsc{HwFuzzEnv} to speedup hardware fuzzing further in an industrial environment.
Since multi-thread techniques do not modify the original single-thread hardware fuzzing flow, we confine our comparison to the speed of hardware fuzzing (measured in the number of tested input clock cycles per second) in the experiments. Table~\ref{tab:multi-thread-speedup} illustrates the speedups achieved by different multi-thread mechanisms (with or without pipeline) in comparison to single-thread.

When the thread count is small, both the simple multi-thread framework and the pipelined framework achieve a speedup roughly equal to the number of threads utilized. However, as the thread count increases, the speedup of the simple multi-thread mechanism does not exhibit linear growth. In this case, the multi-thread pipelined framework demonstrates a noticeable advantage, especially for small designs like I2C or Sodor1Stage. As shown in Table~\ref{tab:multi-thread-speedup}, the multi-thread pipelined fuzzing mechanism with 64 threads achieves a speedup of up to $31.0\times$ (on average $28.2\times$) compared with the single-thread. When compared to \textsc{RtlFuzzLab} without \textsc{HwFuzzEnv} (\ie Original), it achieves a fuzzing speed of up to $621\times$ (on average $361\times$) faster. 
These advancements significantly improve the efficiency of hardware fuzzing for industrial verification.

\begin{table}[t]
\centering
\caption{Speedups of different parallel mechanisms of \textsc{HwFuzzEnv}.}
\vspace{-10pt}
\renewcommand{\arraystretch}{0.85}
\small
\begin{tabular}{|c|c|c|c|c|}
\hline
\textbf{DUT} & \textbf{I2C} & \textbf{Sodor1Stage} & \textbf{Rocket Core} \\ \hline
Original &  0.05$\times$ & 0.06$\times$ & 0.16$\times$ \\ \hline
1 Thread & 1.00$\times$  & 1.00$\times$ & 1.00$\times$ \\ \hline
4 Thread & 3.26$\times$ & 3.19$\times$ & 3.57$\times$ \\ \hline
Pipelined 4 Thread & 3.44$\times$ & 3.34$\times$ & 3.83$\times$ \\ \hline
16 Thread & 9.90$\times$  & 8.85$\times$ & 11.24$\times$ \\ \hline
Pipelined 16 Thread & 12.99$\times$ & 12.45$\times$ & 14.60$\times$ \\ \hline
64 Thread & 15.45$\times$  & 8.94$\times$ & 21.11$\times$ \\ \hline
Pipelined 64 Thread & 31.05$\times$  & 25.79$\times$ & 28.26$\times$ \\ \hline

\end{tabular}
\label{tab:multi-thread-speedup}
\vspace{-5pt}
\end{table}
\setlength{\textfloatsep}{5pt}  

\vspace{-5pt}
\section{Discussion}
As hardware designs scale up and human costs rise, hardware fuzzing emerges as a promising technique for hardware verification automation. This paper aims to advance the integration of hardware fuzzing into industrial verification. Through our analysis and prototype development, we gain a clearer understanding of the current gap between hardware fuzzing and its practical application.

Currently, EDA tools lack sufficient support for hardware fuzzing operations, which substantially limits the efficiency of hardware fuzzing and obstructs its industrial application. For the coverage collection stage, hardware simulators should incorporate a faster mechanism akin to our sketched coverage interface. For the mutation stage, a well-designed mutation engine integrated into the testbench development environment is essential. Additionally, the batch simulation of multiple stimuli is a highly desirable feature for hardware simulators intending to support hardware fuzzing.

For hardware fuzzing to be truly applicable in future hardware verification, the issues mentioned above must be overcome first. Our research could serve as a case study, providing industrial EDA companies with valuable insights to enhance their toolkits.

\vspace{-5pt}
\section{Conclusion}

In this study, we explore the industrial applications of hardware fuzzing. 
We begin by summarizing the standards of compatible hardware fuzzing techniques for industrial verification. 
Then, we analyze the current industrial environment to assess its ability to support these fuzzing methods.
Finally, we design a prototype environment \textsc{HwFuzzEnv} to facilitate hardware fuzzing.
With this environment, the previous hardware fuzzing method achieves a substantial speedup of several hundred times. 
Our work provides beneficial perspectives for EDA companies. By adding only a tiny extra effort, EDA tools can efficiently support hardware fuzzing, encouraging them to consider incorporating these features into their upcoming verification toolkits.

\begin{acks}
This work was partly supported by the National Natural Science Foundation of China (Grant No. 62090021) and the National Key R\&D Program of China (Grant No. 2022YFB4500500).
\end{acks}

\bibliographystyle{ACM-Reference-Format}
\bibliography{long}

\end{CJK*}
\end{document}